 \documentstyle{mn}

 \newcommand{\be}{\begin{equation}}
 \newcommand{\ee}{\end{equation}}
 \newcommand{\beqa}{\begin{equationarray}}
 \newcommand{\eeqa}{\end{equationarray}}

\begin{document}
 \date{}
 \title[Graviation coupling constant]
       {Gravitational coupling constant in higher dimensions}  
 \author[Reza Mansouri and Ali Nayeri]
        {Reza ~Mansouri$^{1, 2}$\thanks{email: mansouri@netware2.ipm.ac.ir}  
         and Ali ~Nayeri$^3$\thanks{email: ali@iucaa.ernet.in}\\
         $^1$Institute for Studies in Theoretical Physics and Mathematics,
             P.O.Box 5746, 19395 Tehran, Iran.\\
         $^2$Department of Physics, Sharif University 
             of Technology, P.O.Box 11365-9161, Tehran, Iran.\\  
         $^3$Inter-University Centre for Astronomy and Astrophysics, Post Bag 4,
             Ganeshkhind, Pune 411007, India}
 \maketitle
 \begin{abstract}
 Assuming the equivalence of FRW-cosmological models and their 
 Newtonian counterparts, we propose using the Gauss law in arbitrary dimension
 a general relation between the Newtonian gravitational constant $G$ and the 
 gravitational coupling constant $\kappa$.
 \end{abstract}

 \section{Introduction}
 It is now a common belief among cosmologists and relativists that although
 spacetime appears smooth, nearly flat, and four dimensional on large scales, at
 sufficiently small distances and early times, it is highly curved with all
 possible topologies and of arbitrary dimensions. The initial idea of 
 Kaluza-Klein has been extensively used in  unified theories of fundamental 
 interactions; e.g. Green, Shwarz, \& Witten \shortcite{gsw}. There the extra   dimensions are assumed to be compactified to 
 Planckian size, and therefore do not display themselves in macroscopic 
 processes. The multidimensional cosmologies based on these ideas have been
 extensively studied in the last years [see e.g. Cho \shortcite{cho}, Kirillov \& Melnikov \shortcite{km}, Wesson \shortcite{wess}]. \\
 Usually in cosmological models based on higher dimensions the problem of  
 the dimensionality of the gravitational coupling constant is not tackled 
 on, being tacitly assumed to be 
 \be
 \kappa = 8 \pi G,  \label{kappa}                                       \ee 
 where  $G$ is the Newtonian gravitational constant. This is ofcourse a relation
 being derived in four dimension. Even in some textbooks this has not been
 differentiated[see e.g. Kolb \& Turner \shortcite{kt}]. 
 The effective gravitational constant in the 
 mutidimensional cosmologies is defined through the multiplication of this 
 four-dimensional value with some volume of the internal space which could 
 even be infinite[Rainer \& Zhuk \shortcite{rz}]. \\  
 There are however cosmological models in higher dimension
 where this picture do not work and one should be cusious about the value of
 the coupling constant in dimensions higher than four[see e.g. Chatterjee \shortcite{ca}, Chaterjee \& Bhui \shortcite{cb}, Khorrami, {\it et. al} \shortcite{kmme}, Khorrami, Mansouri \& Mohazzab \shortcite{kmm}]. In fact, Chatterjee \shortcite{ca}, Chaterjee \& Bhui \shortcite{cb} calculate a homogeneous model in a higher dimensional
 gravity using the same 4-dimensional relation as in (1).\\  
 We propose to show some deficiencies of this misuse and propose a 
 generalization for the gravitational coupling constant for any arbitrary 
 dimension. Any definition of a coupling constant in higher dimension will
 influence the time dependence of $G$ and therfore could have direct 
 observational consequences; e.g. Barrow \shortcite{barr}, Degl'Innocenti
 , {\it et.al} \shortcite{df},  
  Ramero \& Melnikov \shortcite{rm}. \\ 
 In this note we confine ourselves to a mere generalization of the $\kappa$,
 using  discrepancies in comparing relativistic and Newtonian cosmologies 
 in higher dimensions.    
 It belongs to the folklore of the theories of gravitation that their weak 
 limit must be the Newtonian theory of Gravitation. Moreover, on account of the 
 relation (1), there is a Newtonian derivation of the FRW cosmologies. We will
 show that this derivation, using the coupling constant (1) is just valid in 
 $(3+1)$-dimension, and has to be changed for higher diemensional theories.
 On account of the Gauss theorem we give a generalization of it, which makes 
 the Newtonian derivation valid in arbitrary dimensions.

 \section{Friedmann models and Newtonian cosmology in D dimension} 
 
 Consider the following Hilbert-Einstein action in ${\bf D+1}$ dimensional
  space-time with ${\bf D}$ as the fixed dimension of space: 
 
 \be
 S_g = -\frac{1}{2\kappa} \int \ ^{(D+1)}R \sqrt {-g} \ d^{D+1}x
      +\frac{1}{2} \int T \sqrt{-g} \ d^{D+1}x,     \label{h-e action}
 \ee
 
 \noindent where $\kappa$ is the gravitational constant, again. For simplicity,    we consider the $k=0$ FRW model. In an arbitrary fix dimension we ontain  
 the following Friedmann equation, \cite{kmme} 
 \be
 (\frac{\dot a}{a})^2=\frac{2 \kappa}{D(D-1)}\rho, \label{grfr}
 \ee
 which leads to the familiar Friedmann equation in ${\bf D=}3$:  
 \be
 (\frac{\dot a}{a})^2=\frac{\kappa}{3} \rho = \frac{8 \pi G }{3} \rho. 
  \label{grfr=3} 
 \ee
 Now, it is well known that the Friedmann equation (\ref{grfr=3}) can be derived
 from a Newtonian point of view. Taking the Newtonian equation of 
 gravitation in ${\bf D}$ dimension in the usual form  
 \be
 \nabla^2_d \;\varphi = 4\pi G \rho,   \label{dnabla}                                  \ee
 with $\varphi$ the gravitational constant and $\nabla_d$ the ${\bf D}$ 
 dimensional $\nabla$ operator, then the corresponding Friedmann equation 
 can easily be obtained to be  
 \be
 (\frac{\dot a}{a})^2=\frac{8 \pi G }{D(D-2)} \rho.  \label{nfr}
 \ee
 Now, as expected, for ${\bf D}=3$, the familiar relation (\ref{grfr=3})
 is obtained, 
 assuming the relation (\ref{kappa}). But, how if ${\bf D}\neq 3$. Then one will
 realise that the agreement between (\ref{grfr=3}) and (\ref{nfr}) will fail. 
 
 \section{Modification}
 Looking for the roots of the factor $8\pi$ in (1) we come across the relation 
 \be
 R_{00}=\nabla^2 \varphi.  \label{ric-po}
 \ee
 Now, the coefficient in the Poisson equation, i.e. $4 \pi$ has been obtained 
 , using Gauss law, for three dimensional space. Thus we should first derive  
 the correct coefficient for a {\bf D} dimensinal space. Applying  
 Guass's law for a {\bf D} dimensional volume, we find the Poisson equation 
 for arbitrary fixed dimension,
 \be
 \nabla^2 \varphi = \frac{2 \pi ^{D/2}  G}{(D/2-1)!} \rho. \label{modpoiss}
 \ee
 On the other hand we get for arbitrary $D$
 \be
 R_{00}=(\frac{D-2}{D-1}) \kappa \rho. \label{ricc-kappa}
 \ee
 A comparison of (\ref{ric-po}), (\ref{modpoiss}), and (\ref{ricc-kappa}) will
 give us the following modified Einstein gravitational constant,
 \be
 \kappa_D = \frac{2(D-1)\pi^{D/2}  G}{(D-2)(D/2-1)!}. \label{modkappa}
 \ee
 The correct form of Friedmann equation in any 
 arbitrary fixed dimension is now derived to be 
 \be
 (\frac{\dot a}{a})^2 = \frac{4 \pi^{D/2}  G}{D(D-2)(D/2-1)!}\rho.  
 \ee
 As it is easily seen, the above relation is in complete agreement with its
 Newtonian counterpart in all dimensions.
 
 \section{conclusion}
 The dimensional dependence of the gravitational constant $\kappa$ may have 
 very different and serious field theoretic and astrophysical consequences
 hitherto unnoticed. 
 It would be interesting, e.g., to see which changes are to be expected if
 the results of this note are combined with Kaluza-Klein paradigm. The 
 consequences on time variation of G within various models is another
 very interesting cosmological issue which is under investigation.   
 
 \section*{ACKNOWLEDGEMENT}
 A.N. wishes to thank Prof.Padmanabhan for many useful discussions. A.N. was financially supported by the Council of Scientific and Industerial Research, India.

 \end{document}